\documentclass[
 reprint,
 amsmath,amssymb,
 aps,
 floatfix,
]{revtex4-2}
\usepackage[utf8]{inputenc}
\usepackage[T1]{fontenc}
\usepackage[english]{babel}
\usepackage{graphicx}
\usepackage{dcolumn}
\usepackage{bm}
\usepackage{float}
\usepackage{siunitx}
\usepackage{hyperref}
\renewcommand{\d}[1]{\mathop{}\!\mathrm{d}#1\!\mathop{}}

\newcommand{\armchair}{_a}
\newcommand{\zigzag}{_z}

\newcommand{\wext}{w_\text{ext}}

\newcommand{\crit}{{\scriptscriptstyle (}crit{\scriptscriptstyle )}}
\renewcommand{\matrix}{\mathbf}

\newcommand{\smallN}{32}
\newcommand{\smallw}{12}
\newcommand{\bigN}{80}

\usepackage{subfigure}
\hypersetup{
colorlinks=true,
linkcolor=blue,
filecolor=blue,
urlcolor=blue,
citecolor=blue,
}
\usepackage{orcidlink}
\begin{document}
\preprint{APS/123-QED}
\title{Charge carrier flow through trimmed graphene nanoribbon junctions}
\author{Julien Leuenberger\,\orcidlink{0009-0000-0701-6097}}
\author{Kristi\=ans \v Cer\c nevi\v cs\,\orcidlink{0000-0001-8373-5456}}
\author{Oleg V. Yazyev\,\orcidlink{0000-0001-7281-3199}}
\email{oleg.yazyev@epfl.ch}
\affiliation{Institute of Physics, Ecole Polytechnique F\'ed\'erale de Lausanne (EPFL), CH-1015 Lausanne, Switzerland}
\date{\today}
\begin{abstract}
As Moore's law approaches its fundamental limits, the development of nanoelectronic devices using low-dimension materials has become a promising avenue for further miniaturization and performance improvements. Among the various novel materials, graphene nanoribbons (GNRs) have emerged as particularly attractive candidates due to their unique electronic properties, opening up a whole new nanoelectronics paradigm consisting of circuits made entirely of graphene. However, due to the technical constraints that naturally arise when working on a two-dimensional plane, the design of efficient nanoelectronic components with a minimal spatial footprint remains a significant challenge. In particular, connecting various components can be a real architectural challenge, comparable to that of the first printed circuit boards. This paper investigates strategies for designing optimal-sized nanoribbon junctions which allow connecting GNRs at an angle, by trimming the junction edge while maintaining favorable electronic properties. Specifically, we show that the probability density current at the tip of junctions is negligible, implying that a selection of atoms can safely be removed without significantly altering the conductance. More generally, we demonstrate that larger trimmings have impacts on the conductance channels, resulting in a conductance that is mainly dictated by the ratio of armchair and zigzag edges. Finally, we propose a simple model relating this ratio to the conductance.
\end{abstract}
\maketitle
\section{Introduction}
Moore's Law, proposed by Intel co-founder Gordon Moore in 1965, predicted that the number of transistors on a microchip would double approximately every two years, leading to an exponential increase in computing power \cite{Moor65, Moor75}. For decades, this prediction held, driving remarkable technological progress and enabling the development of the modern digital era \cite{Mack11}. However, in recent years, there have been growing concerns that the physical limitations of existing semiconductor technology are beginning to impede further progress, with many experts predicting an end to the historical trend \cite{4567410, Thei17}. This has spurred intensive research efforts aimed at overcoming these limitations and finding new avenues for continued advancement in the field of computing.
One promising area of focus is low-dimensional materials, which exhibit unique physical properties \cite{Novo04, Wang12, Iann18} that could potentially overcome the limitations of traditional 3D nanoelectronics. Among these, graphene and graphene nanoribbons (GNRs) have emerged as up-and-coming options \cite{Ares07, Chen07, Avou10, Wang2021, Rads21}. GNRs are narrow strips of graphene just a few atoms wide, and their electronic properties can be tuned by adjusting their width, edge structure, topology, and doping \cite{Naka96, Son06a, Baro06, Yazy13, Chen13, Gron18, Pizz22, RevModPhys.81.109}. In 2007, Areshkin and White laid the foundations for a new paradigm in nanoelectronics by proposing innovative components based on GNRs of various size and shapes, making it possible for the first time to envisage electronic circuits composed solely of graphene \cite{Ares07}. While the main difficulty in developing such circuits was the very high precision required, Cai \emph{et al.} demonstrated in 2010 a novel method for obtaining GNRs with atomically precise edges using a combination of bottom-up synthesis and surface-assisted assembly \cite{Cai10}. Whereas  top-down methods were only able to achieve precision in the order of ten nanometers \cite{Han07, Chen07, Li08, Datt08}, this  breakthrough has opened up new possibilities for designing and engineering GNRs with tailored electronic and magnetic properties \cite{Cai14, Llin17, Rizz18, Rizz21, Blac21, Majh22, Bori22}.
While the potential of two dimensional electronic devices is exciting, there are still significant hurdles that must be overcome to realize their full potential. One of the biggest challenges is designing and fabricating of effective interconnects, which are essential for connecting the various components of such chips and circuits. Since in a two-dimensional space wires can't overlap or cross, finding efficient layouts is a crucial role in the elaboration of future circuits, where only turns and bends remain as a design option. However, due to the ballistic transport regimen at this scale, even minor changes in the structure can have a significant impact on their properties \cite{Cern23}. We thus need to obtain a more clear insight into the relation between the exact shape of these angled junctions and their electronic properties. While there are numerous proofs that the edge structure plays a crucial role in the system's conductance \cite{Naka96, Son06a, Baro06, Chen17a, Pizzochero_2021}, these results still do not produce overarching guidelines for junction engineering. On the other hand, the study of the overall shape's influence on the transport properties has led to interesting models -- such as the so-called graphical atomic orbital scheme \cite{Zhao16, Tsuji18} -- but which do often only consider systems where each lead is connected to the considered structure at only a single point.
We decided to pursue the investigation further, by studying the effect of shape, size, and edge structure on the transport properties of large angled GNR junctions, and most notably on their conductance. A more in-depth analysis will then be made to understand these results, mainly by inspecting other electronic properties of the junctions, including the local density of states (LDOS) and the spatial distribution of the currents. This will help achieve a better insight into the relationship between geometry and electronic properties. Finally, we also develop a predictive model to optimize the geometry of arbitrary GNR junctions.
\section{Method}
\begin{figure}[t]
\includegraphics[width=\columnwidth]{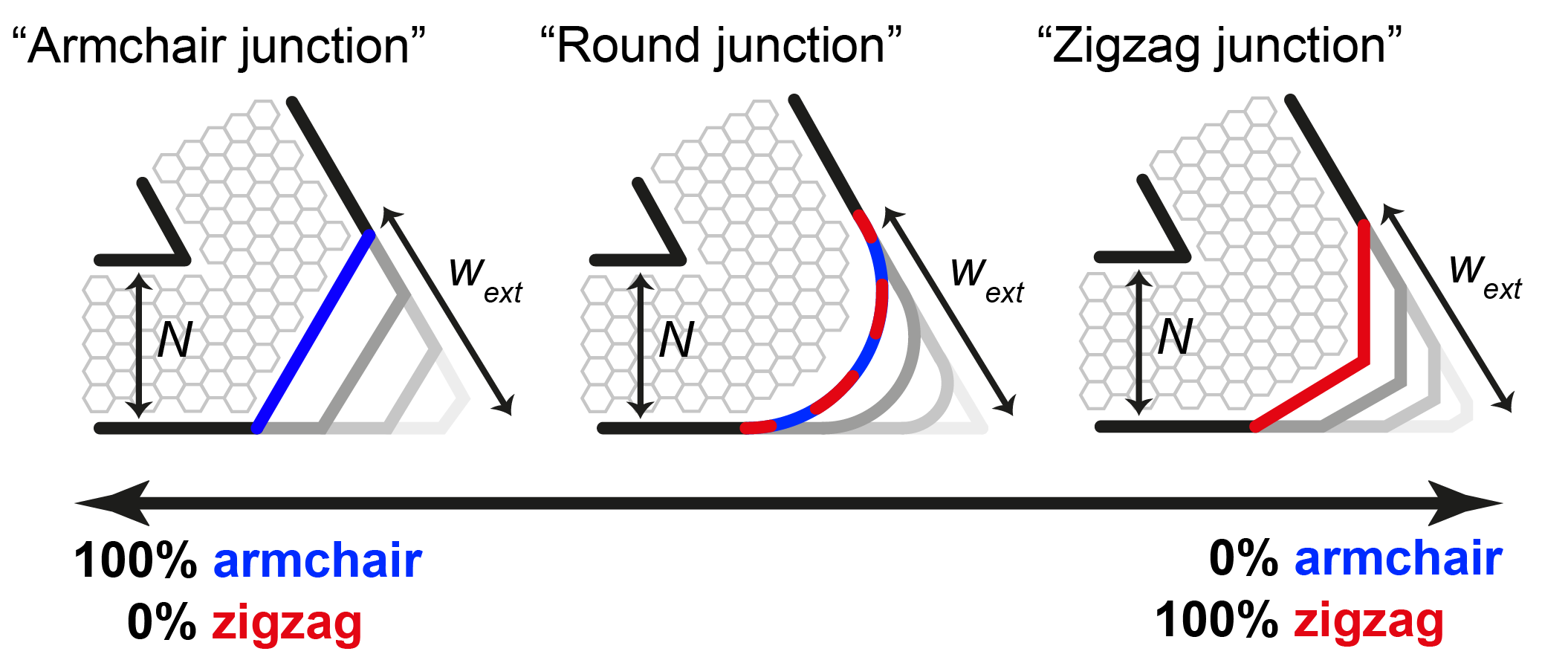}
\caption{Schematic view of the different $N$-AGNR junctions. From left to right: an \emph{armchair junction}, a \emph{round junction}, and a \emph{zigzag junction}. Newly defined edges are colored. Blue and red edges correspond to armchair and zigzag edges respectively. Leads' edges are in black. The gray lines at the tip of the junctions represent the trimming process, quantified by the cutoff parameter $\wext$.}
\label{schematic junctions}
\end{figure}

Tight-binding (TB) model with one $p_{z}$ orbital per atom and only first nearest-neighbor (1NN) hopping integral are used to investigate the electronic properties of GNR junctions. This semi-empirical method has been shown to provide a computationally inexpensive, but accurate characterization of the electronic properties of graphene \cite{Kund09a} and GNRs \cite{Wimm08, Hanc10} near the Fermi level, enabling the systematic study of a wide range of junctions which in turn will make it possible to investigate certain topology-related trends.
Making use of the second quantization, the well-known Hamiltonian for a given device $\hat{H}_D$ reads
\begin{equation}\label{eq:hamiltonian}
\hat{H}_{D}=\sum_{j}\varepsilon_0\hat{c}_{j}^{\dagger}\hat{c}_{j}-t\sum_{j,k}(\hat{c}_{j}^{\dagger}\hat{c}_{k}+\text{h.c.}),
\end{equation}
where $t=2.75$~eV is the nearest-neighbour hopping integral, $\varepsilon_0=0$~eV is the on-site energy, $\hat{c}_{\mu}^{\dagger}(\hat{c}_{\mu})$ creates (annihilates) an electron on site $\mu$, and the sum $(j,k)$ is restricted to nearest neighbor atoms. In the case of circuits made entirely of graphene, values of $t$ and $\varepsilon_0$ are the same throughout the entire system.
Transport properties are calculated through Green's function formalism which, after discretizing the spatial coordinates, is given by
\begin{equation}
\matrix{G}_{f}(E)=\left((E+i\eta) \matrix{I}-\matrix{H}_{D}-\matrix{\Sigma}_{L}(E)-\matrix{\Sigma}_{R}(E)\right)^{-1},
\label{eq:green}
\end{equation}
where $\eta$ adds an infinitesimally small imaginary character to energy $E$, $\matrix{I}$ is the identity matrix and $\matrix{\Sigma}_{L(R)}$ is the self-energy that is obtained self-consistently using
\begin{equation}
\matrix{\Sigma}_{L(R)}(E)=\matrix{H}_{1}^{\dagger}(E\matrix{I}-\matrix{H}_{0}-\matrix{\Sigma}_{L(R)}(E))^{-1}\matrix{H}_{1},
\label{eq:dyson}
\end{equation}
$\matrix{H}_{0}$ being the Hamiltonian of the lead unit cell and $\matrix{H}_{1}$ being the coupling. The broadening function $\matrix{\Gamma}_{L(R)}$ due to the coupling with the leads is calculated from the self-energies:
\begin{equation}
\matrix{\Gamma}_{L(R)}(E)=i[\matrix{\Sigma}_{L(R)}(E)-\matrix{\Sigma}_{L(R)}(E)^{\dagger}].
\label{eq:gamma}
\end{equation}
The transmission coefficient $T$ can be expressed as
\begin{equation}
T(E)=\mathrm{Tr}[\matrix{\Gamma_{L}}\matrix{G}_{f}\matrix{\Gamma}_{R}\matrix{G}_{f}^{\dagger}].
\label{eq.trans}
\end{equation}
Finally, we express conductance $G$ in terms of conductance quantum $G_{0}$ using the Landauer formula \cite{Land57a}
\begin{equation}
G(E)=G_{0}T(E)=\dfrac{2e^{2}}{h}T(E).
\end{equation}
Green's function formalism allows the computation of various quantities of interest. The local density of states (LDOS) at site $j$ is given by
\begin{equation}
    \text{LDOS}_j(E)=-\frac{1}{\pi}\text{Im}\left([\matrix{G}_f(E)]_{jj}\right)
\end{equation}
Note that all numerical calculations are carried out using Python's package \textsc{kwant} \cite{kwant}, which uses a wave function approach based on the scattering matrix formalism, equivalent to the Green's function formalism thanks to the Fisher-Lee relation \cite{Fish81}. This allows to write the local current flowing from site $j$ to $k$ as
\begin{equation}
    J_{kj}=i(\psi_j^\dagger [\matrix{H}_D]_{kj}\psi_k-\psi_k^\dagger [\matrix{H}_D]_{kj}\psi_j),
\end{equation}
where $\psi_\mu$ is the value of the wave function at site $\mu$.
In order to quantify the efficiency of a given junction with respect to the pristine leads, we introduce the descriptor of \emph{preserved conductance} $\tau$ \cite{Pizzochero_2021}, as
\begin{equation}\label{tau_eq}
\tau(E)=\frac{\int_{E-\frac{1}{2}\delta E}^{E+\frac{1}{2}\delta E}G(E')\d{E'}}{\int_{E-\frac{1}{2}\delta E}^{E+\frac{1}{2}\delta E}G_\text{lead}(E')\d{E'}},
\end{equation}
where $G(E)$ and $G_\text{lead}(E)$ are the conductance of the junction respectively the lead at energy $E$, and $\delta{E}$ is extending over a small energy window. In this paper, we consider the transport properties around the Fermi level as it is the most relevant energy range for interconnects.
One of the simplest ways of lowering the footprint of the angled junction is by trimming its tip. We can control the amount of removed atoms by introducing a cutoff parameter $\wext$. To give it a clear definition, let's consider an equilateral triangle of side length $\wext$ (in units of the leads' periodicity $\sqrt{3}a$, with $a$ being the lattice parameter of graphene). All atoms lying inside this triangle-like shape are then removed. Therefore, the larger $\wext$, the more atoms are removed and the longer the newly defined edge. More generally, we can also consider other cut-off paths. For instance, atoms can be removed following a circular arc (thus being of radius $R=\frac{1}{3}\wext$). Since it is well known that the edge structure of graphene has an impact on the conductance \cite{Naka96, doi:10.1143/JPSJ.65.1920}, the effect of such trimmings will be studied by comparing three of them: one whose cutoff path is circular, one which maximizes the number of armchair edge atoms, and one which maximizes the number of zigzag edge atoms. We will hence refer to these junctions as \emph{round}, \emph{armchair} and \emph{zigzag} junctions. We show the considered families and the evolution of such GNR junctions in Fig.~\ref{schematic junctions}.
In the scope of this study, we will focus our attention to leads whose width is $N=3n+2$ which are metallic in the 1NN approximation, and whose bandgap is much smaller compared to the $N=3n$ and $N=3n+1$ families in practice \cite{Yama20}. Furthermore, we consider leads of large width (namely $20 \leq N \leq 80$) since they tend to have lower bandgaps, more suited for interconnects \cite{Son06a, Yang07, Zhu11}.

\section{Results \& discussion}
\begin{figure*}[t]
\includegraphics[width=\textwidth]{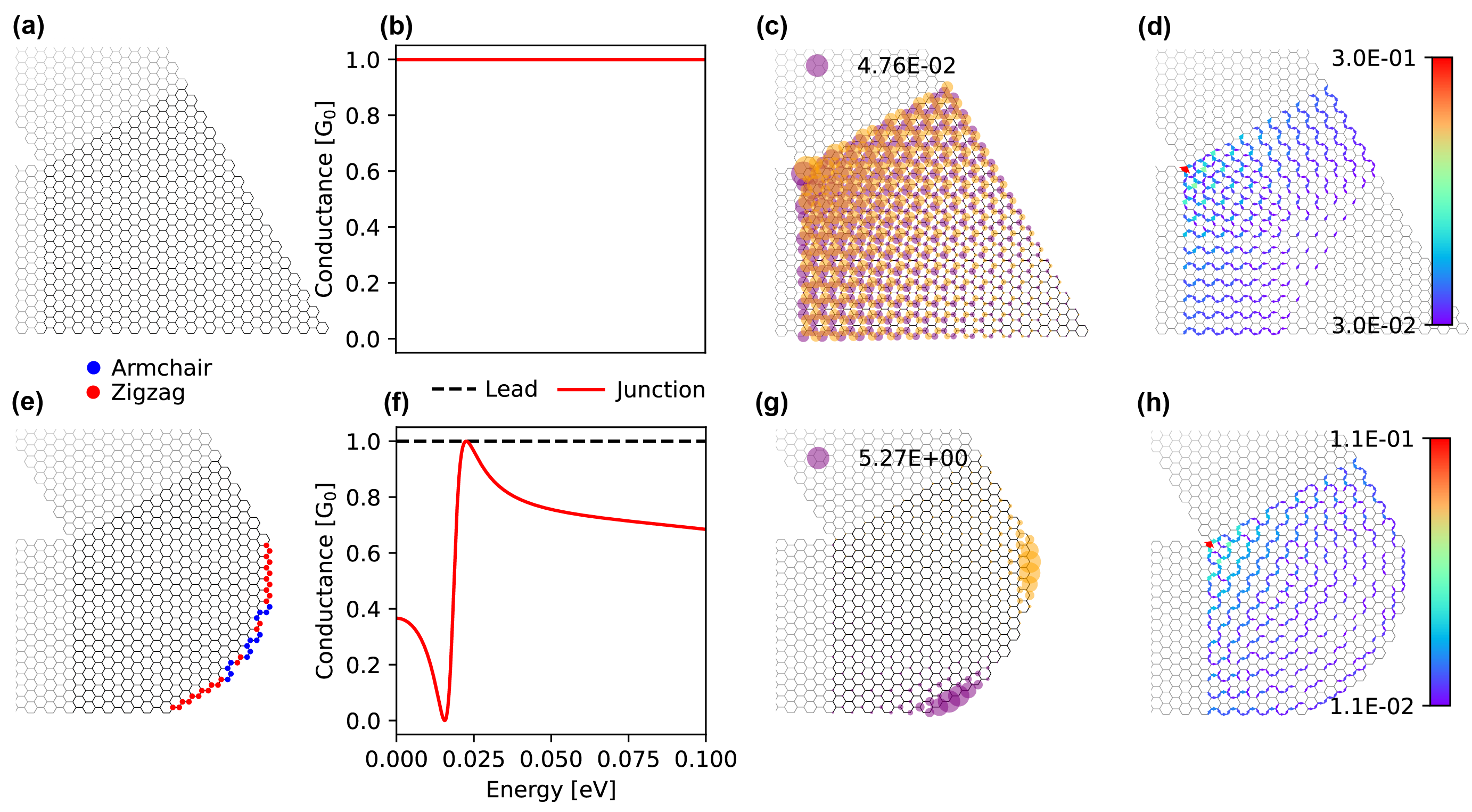}
\caption{Physical and electric properties of two types of $32$-AGNR junctions. (a) Edge structure of the so-called sharp junction. (b) Conductance of the sharp junction (red solid line) and of one of the leads (black dashed line). (c) Local density of states (LDOS) of the sharp junction at the Fermi level. Colors indicate the corresponding sublattice. (d) Probability current of the sharp junction at the Fermi level. For readability purposes, the lowest values are not displayed. (e) Edge structure of the so-called round junction, with a rounding of $\wext=32$. Colored atoms correspond to the newly defined segment. The blue and red coloring correspond respectively to armchair and zigzag edges, as defined in Fig. S1. (f)--(h) are the round junction's conductance, LDOS, and probability current.}
\label{fig:round_junctions}
\end{figure*}
\begin{figure*}[t]
\includegraphics[width=\textwidth]{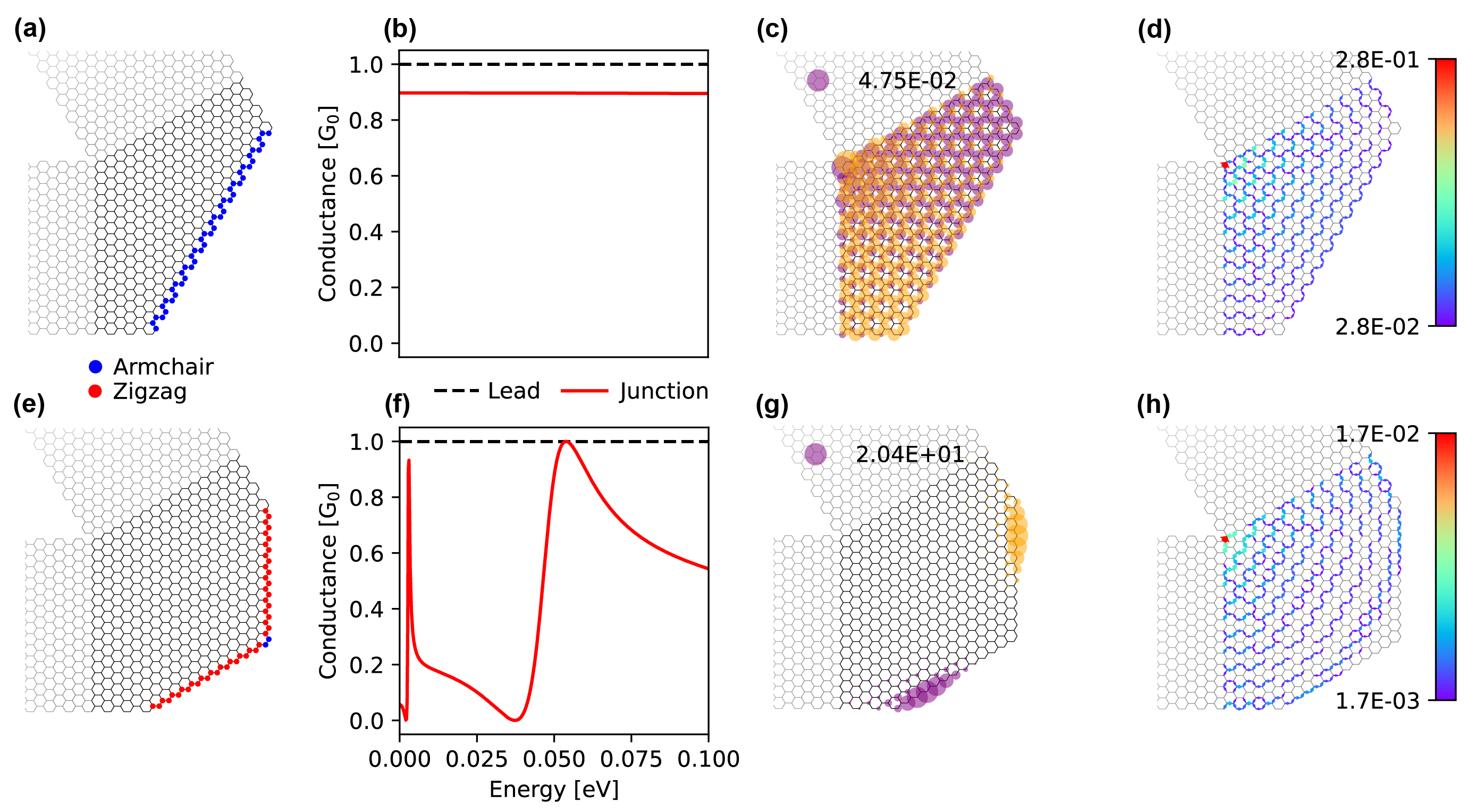}
\caption{Physical and electric properties of $\smallN$-AGNR junctions with $w_\text{ext}=\smallw$. (a) Edge structure of the armchair junction. The blue dots represent local armchairs while the red ones represent local zigzags. (b) Edge structure of the armchair junction. The dots correspond to the different types of edges. (c) Local density of states (LDOS) of the armchair junction at the Fermi level. Colors indicate the corresponding sublattice (d) Probability current of the armchair junction at the Fermi level. For readability purposes, the lowest values are not displayed. (e)--(h) correspond to the same data but for the corresponding zigzag junction.}
\label{fig:extreme_junctions}
\end{figure*}
\begin{figure*}[t]
\includegraphics[width=\textwidth]{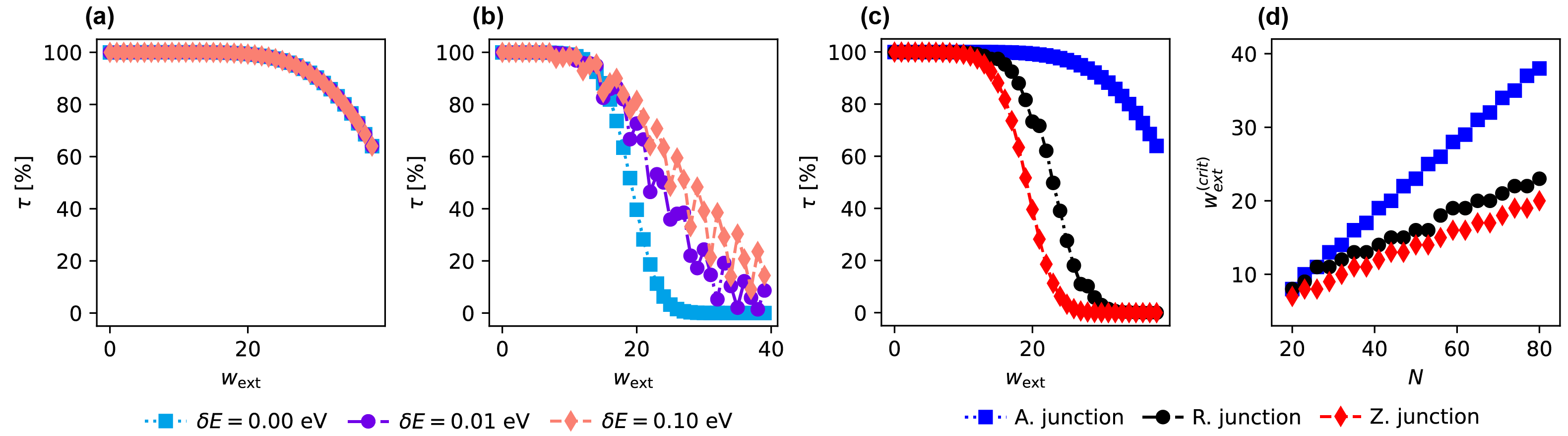}
\caption{Preserved conductance as a function of the cutoff parameter $\wext$ for energy windows of $\delta E=\qty{0.00}{\eV}$ (blue squares), $\delta E=\qty{0.01}{\eV}$ (purple dots) and $\delta E=\qty{0.10}{\eV}$ (red diamonds), for $\bigN$-AGNR (a) armchair junctions, and (b) zigzag junctions. (c) Comparison of the preserved conductance at $\delta E=0$ of armchair junctions (blue squares), round junctions (black circles), and zigzag junctions (red diamonds). (d) Largest cutoff parameter $\wext^{\crit}$ leading to a preserved conductance of $\tau=90\%$ for armchair junction (blue squares), round junctions (black circles), and zigzag junctions (red diamonds).}
\label{tauvsr}
\end{figure*}

We begin our investigation by examining a perfectly conducting ($\tau=100\%$) \ang{60} 32-AGNR junction [see Fig.~\ref{fig:round_junctions}(a)--(d)] obtained by fully extending the two leads until the intersection point in order to obtain a pure armchair edge junction. Due to the pointed protruding scattering region, we will refer to this junction as a \emph{sharp junction}. It has been shown before \cite{Chen17a, Cern23} that preserving the lead geometry and edge type greatly improves the electronic transport properties of the junction. Therefore such a junction already would exhibit promise as an interconnect. However, their relatively large size could hamper their implementation in future compact circuits. We will therefore explore ways of reducing their footprint while maintaining their transport properties. It can be observed in Fig.~\ref{fig:round_junctions}(c) and \ref{fig:round_junctions}(d) that while we can notice electron delocalization over most of the junction, there is a preference for stronger probability current on the inside edge. This finding indicates that geometry modifications on the outer edge would exhibit a smaller effect and thus allow reducing the size of the junction.
Therefore, in our first attempt to optimize the size of the junction, we trim the sharp angle on the outside of the junction to obtain a rounded end. An example junction with the cutoff parameter $\wext=12$ is displayed in Fig.~\ref{fig:round_junctions}(e). Note that more examples can be seen in Figs. S4 and S5 of the Supplementary Information (SI) as well as online \cite{Leu23}. We can notice that the scattering region now contains a mix of armchair and zigzag edges. This, in turn, leads to significant changes in the conductance plots in Fig.~\ref{fig:round_junctions}(f), where pronounced Fano anti-resonances due to the localized states can be observed \cite{Miro10}. Comparing the probability current maps between the sharp [Fig.~\ref{fig:round_junctions}(d)] and rounded [Fig.~\ref{fig:round_junctions}(h)] junction, we do notice changes in the pathways as well as in intensities. While the sharp junction exhibits conductance channels similar to the ones of pristine AGNRs \cite{Waka99a}, the probability current in the round junction has a less well-defined macroscopic structure. The LDOS plots in Fig.~\ref{fig:round_junctions}(c) and \ref{fig:round_junctions}(g) display qualitatively differing characters. It can be seen that, at $E=0$, significant localization occurs on the rounded part of the junction due to the zero-energy states associated with zigzag edges \cite{Naka96, Waka99a}. Further evidence of the zigzag-edge states arises from the fact that these states are only associated with one of the sublattices as seen by the differently colored sublattice LDOS in Fig.~\ref{fig:round_junctions}(g).
\begin{figure}[t]
\centering
\includegraphics[width=\columnwidth]{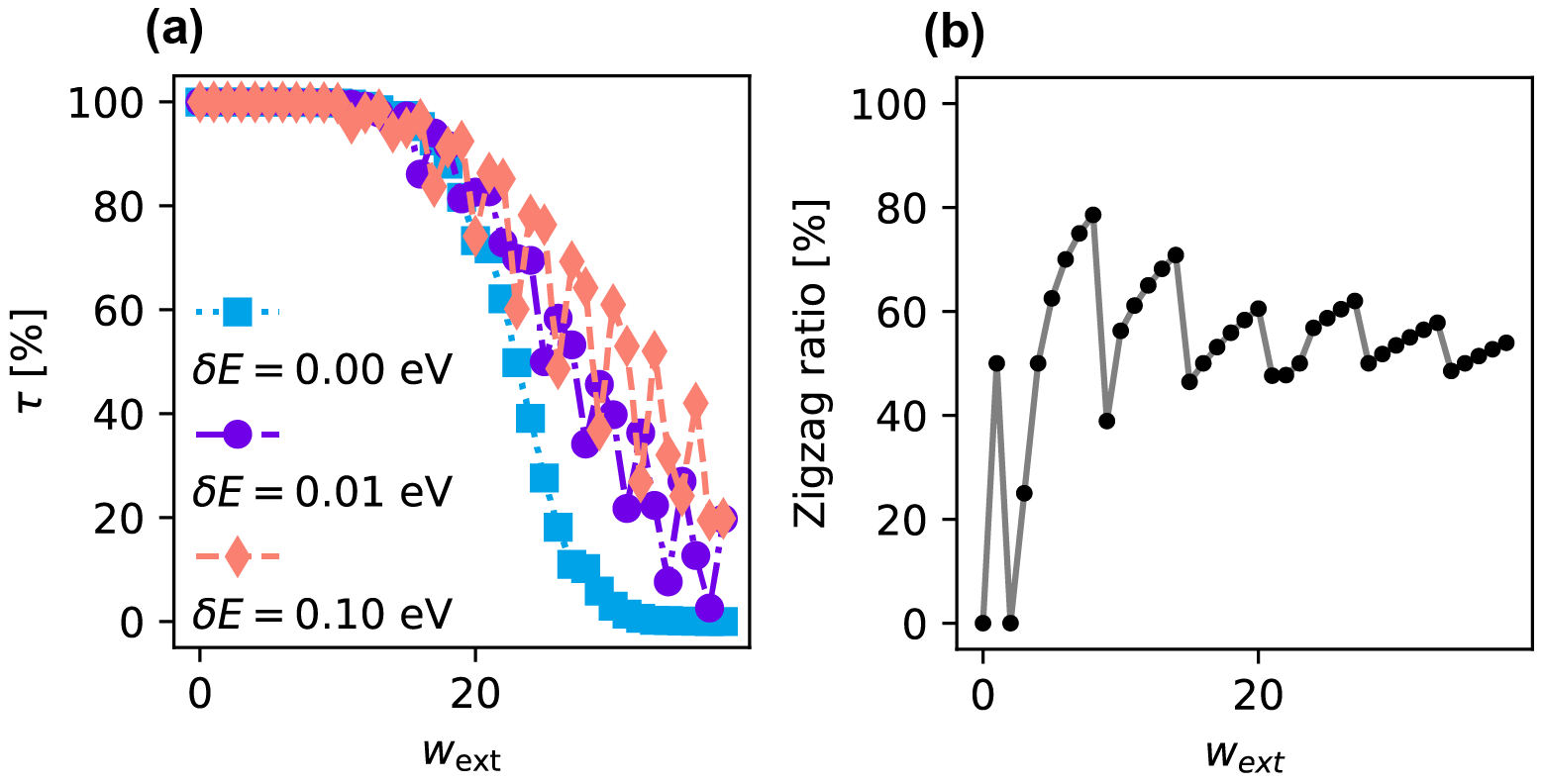}
\caption{(a) Preserved conductance as a function of the cutoff parameter $\wext$ of $\bigN$-AGNR junctions, for energy windows of $\delta E=\qty{0.00}{\eV}$ (blue squares), $\delta E=\qty{0.01}{\eV}$ (purple dots) and $\delta E=\qty{0.10}{\eV}$ (red diamonds). (b) The ratio of armchair edges. Lines are for readability.}
\label{tauvsr_round_ratio}
\end{figure}
To systematically investigate the rounding effect, we calculate the transport properties for various cutoffs in the vicinity of the Fermi level and display the results for 80-AGNR junctions in Fig.~\ref{tauvsr_round_ratio}. We notice that for small cutoff values ($\wext\lesssim20$) the perfect ($\tau\approx100\%$) conductance in the vicinity of the Fermi level is preserved. However, a sharp phase transition and then complete conductance suppression can be observed for $\wext\gtrsim25$. This effect is more pronounced for $\delta E=0$, whereas other $\delta E$ values induce pronounced oscillating behavior and a more gradual decrease in preserved conductance. Based on these observations, we establish that small changes to the atomic structure of the outer edge have a minor influence on the conductance, mainly because of lower interaction with the conductance channels. Hence, more optimal junction sizes can be engineered without losing the beneficial transport properties.
As the trimming of the junctions incurs changes in the edge type composition [see Fig.~\ref{tauvsr_round_ratio}(b)], we investigate the effects on conductance by considering two bounding cases -- pure armchair or pure zigzag junctions as discussed in Fig.~\ref{schematic junctions}. Therefore, we examine junctions with similar parameters as the round one in Fig.~\ref{fig:round_junctions}, namely 32-AGNR with $\wext=12$. Their physical and electronic characteristics are presented in Fig.~\ref{fig:extreme_junctions}. Starting with Fig.~\ref{fig:extreme_junctions}(b) we notice constant conductance of $0.9\ G_{0}$ across $|E|< 0.1$ eV for the armchair junction, however significant Fano anti-resonances can be observed in the case of zigzag junction in Fig.~\ref{fig:extreme_junctions}(f). Similar behavior was observed in the case of rounded junction in Fig.~\ref{fig:round_junctions}(f) hence revealing the major role played by the zigzag edge segments in our junctions. Further evidence of the zigzag edge effect can be recognized by investigating the LDOS patterns in Fig.~\ref{fig:extreme_junctions}(c) and \ref{fig:extreme_junctions}(g), where strong localization on the individual sublattices can once again be observed in pure zigzag junctions, whereas completely delocalized states in the case of the armchair junction. Finally, Fig.~\ref{fig:extreme_junctions}(d) shows how the armchair junction exhibits the typical AGNR current channels similarly to the previously discussed sharp junction [Fig.~\ref{fig:round_junctions}(d)] while the zigzag junction's currents distribution in Fig.~\ref{fig:extreme_junctions}(h) is less structured.
In general, pure armchair junctions preserve conductance even for large $\wext$ values, as shown in Fig.~\ref{tauvsr}(a). This behavior is consistent across all studied pure armchair junctions and also results in negligible variations of the preserved conductance with increasing energy window, as there are no localized states near the Fermi level. Astonishingly we can reduce the size of the scattering region by almost 50\% and still preserve virtually perfect conductance. On the other hand, pure zigzag junctions display the characteristic phase transition and completely suppressed conductance for $\delta E=0$ with large $\wext$ values, as shown in Fig.~\ref{tauvsr}(b). Note, that in this case there is also an apparent variation in $\tau$ values depending on the chosen $\delta E$. Since rounded junctions also display pronounced conductance dips near the Fermi level and comparable $\tau$ vs $\wext$ plot [see Fig.~\ref{tauvsr_round_ratio}(a)], they give us an indication of the significant impact of zigzag edge segments on the overall conductance properties of graphene junctions even at sparse occurrences \cite{Naka96}.
Next, comparing the evolution of preserved conductance against increasing $\wext$ of the pure armchair and zigzag junctions with those of the rounded junctions at $\delta E=0$, reveals that they respectively constitute upper and lower bounds for any arbitrary junction, as illustrated for $80$-AGNR in Fig.~\ref{tauvsr}(c). Remarkably, this property exhibits a high degree of generality and predictability for all $N$-AGNR junctions. All three junction types exhibit perfect conductance of $\tau \approx 100 \%$ up to $\wext=16$ indicating that the protruding outer edge of the junction does not play a major role in the electron transport regardless of the outer edge composition.
We attribute this property to the previously observed probability current pathways grouped at the inner edge and center of the junctions as well as the upper limit of distinct paths based on the width of the leads. Therefore increasing the size of the scattering center towards the outer edge even further does not unlock additional current paths nor increase conductance. However, at smaller sizes further trimming of the junction displays differing behavior between the three types, hence unlocking the tunability of the junction edge type and ensuing transport properties over the full range of $\tau$.
Furthermore, to enhance our size optimization methodology, we define a critical value $\wext^{\crit}$ as the maximal value of $\wext$ for which the preserved conductance remains above a given threshold. Fig.~\ref{tauvsr}(d) shows the values of $\wext^{\crit}$ for a threshold of $\tau =90\%$ for various $N$-AGNR junctions. Interestingly, we see a clear linear dependence between the critical trimming $\wext^{\crit}$ and the width of the junctions, implying that the decrease in $\tau$ does not depend solely on the amount of trimming $\wext$, but rather on the ratio between $\wext$ and $N$. This allows us to establish a simple design rule governing the size of the junction. Note the clear difference between armchair junctions on the one hand, and zigzag and round junctions on the other. This discrepancy arises from the complete absence of zigzag vertices in armchair junctions, enabling them to be further trimmed before conductance has fallen below the previously set threshold. By following the linear relationship, we can find the optimal size of the junction that preserves at least 90$\%$ of the conductance at $E=0$ for all three junction types. This finding is a direct consequence of the spatial distribution of the probability currents, where the inner edge dominates the transport.
Although we find comprehensive design rules for junction size optimization based on the size and edge type, the exact relationship between the preserved conductance of an arbitrarily trimmed junction and its pure counterparts is still to be determined. Since the presence and amount of zigzag edges play a strong role in the preserved conductance, we introduce the ratio of armchair and zigzag vertices $r_a$ and $r_z=1-r_a$ to characterize our round junctions. A similar approach has already been proposed by Nakada and Fujita back in 1996, demonstrating the great interest this metric represented \cite{Naka96}. An early assumption would be that, by increasing the number or size of zigzag edge segments, we would approach the conductance characteristics of a pure zigzag junction.
We can thus make an Ansatz by using a weighted average between the preserved conductance of the corresponding armchair and zigzag junction, $G\armchair(E)$ and $G\zigzag(E)$ respectively (i.e. for junctions with the same $N$ and $\wext$ parameters as the considered trimmed junction). To determine the most appropriate average to choose, note that the conductance of the round junction goes to zero when the zigzag does the same [see Fig.~\ref{tauvsr}(c)]. In order to recover this property, we'll have to make use of the geometric mean, being
\begin{equation}\label{Ansatz 2}
G(E)\overset{\text{Ansatz}}{=}G\armchair(E)^{r_a} \cdot G\zigzag(E)^{r_z}.
\end{equation}
This Ansatz allows for predictions that are consistent with the calculated conductance trends for all values of $\wext$. Note that the quality of the predictions only marginally depends on the width $N$ of the junctions, with results being only slightly better for large $N$ values due to the longer scattering region segments and thus more precise and slower varying edge type ratios. Further details regarding the quality of the predictions can be found in the SI, notably in Figs. S2 and S3.
\begin{figure}[t]
\includegraphics[width=\columnwidth]{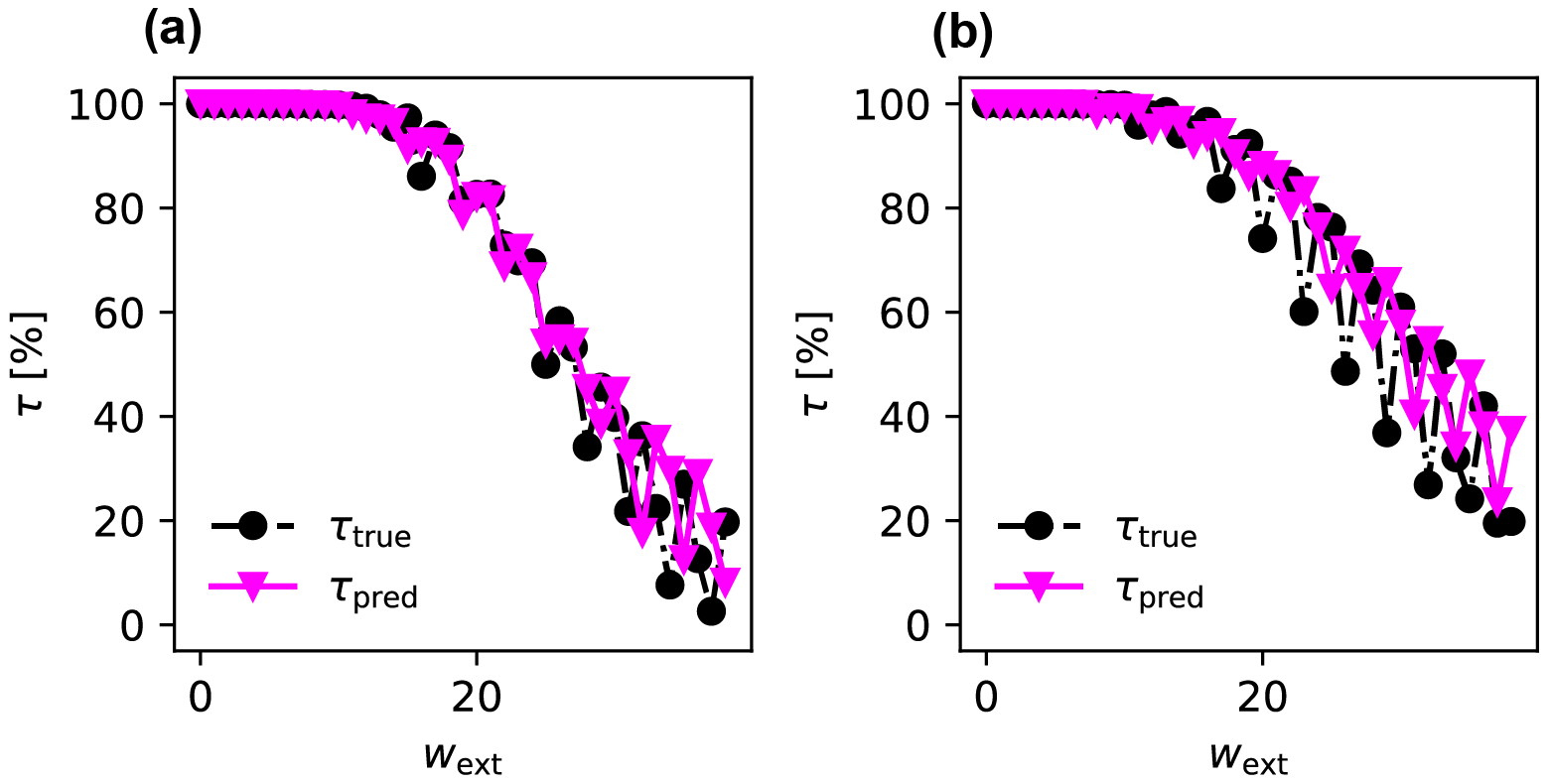}
\caption{Preserved conductances $\tau$ (black dots) and predictions (magenta triangles) as functions of the cutoff parameter $\wext$ using Eq. \eqref{Ansatz 2}, for (a) $\delta E=\qty{0.01}{\eV}$ and (b) $\delta E=\qty{0.10}{\eV}$. Lines are for readability.}
\label{taubounded_E>0}
\end{figure}
Taking into account energy windows $\delta E>0$ around the Fermi level, results become less trivial due to the presence of conductance dips near the Fermi level. Typically, the preserved conductance of a given round junction is not compelled to be bounded anymore by the corresponding armchair and zigzag junctions. Indeed, there are energies for which armchair junctions are performing worse than round or zigzag junctions (e.g. comparing Figs.~\ref{fig:round_junctions}(f), \ref{fig:extreme_junctions}(b) and \ref{fig:extreme_junctions}(f) at $E\sim 0.60$ eV). However, our model still predicts the overall trend of the decay as shown in Fig.~\ref{taubounded_E>0}, and matches the critical $\wext$ value of the phase transition. The quality of these predictions is particularly remarkable. It is important to remember that the Ansatz used to make these predictions is simplified and does not precisely predict the actual value of conductance of a junction at a specific energy (apart at and around the Fermi level, as demonstrated), and thus does not predict phenomenons like anti-resonances.
\section{Conclusion}
In conclusion, our study has demonstrated design guidelines for optimizing the size of angled graphene nanoribbon junctions for applications in nanoelectronic devices. Through systematic analysis of wide graphene nanoribbon junctions, we have identified several critical factors that impact the conductance of a junction and developed a predictive model that can evaluate this behavior with reasonable accuracy. We discover that due to the spatial nature of the probability currents, junctions can be substantially reduced in size by removing atoms near the outside edge without significantly diminishing the transport properties. Moreover, our findings suggest that keeping the number of zigzag edge segments as low as possible is another key factor in maintaining high conductance close to the Fermi level. Further refinements and improvements, such as the inclusion of spin in our model, could have important applications in nanoelectronics. The results of our study represent an important step forward in meeting the challenges posed by the limits of Moore's Law, limits which may only be overcome by the advent of new paradigms in the field of nanoelectronics, such as all-graphene circuits.
\bibliography{references}
\end{document}


\preprint{APS/123-QED}
\title{Supplementary Information}
\author{Julien Leuenberger}
\author{Kristi\=ans \v Cer\c nevi\v cs}
\author{Oleg V. Yazyev}
\email{oleg.yazyev@epfl.ch}
\affiliation{Institute of Physics, Ecole Polytechnique F\'ed\'erale de Lausanne (EPFL), CH-1015 Lausanne, Switzerland}
\date{\today}
\maketitle
\section*{Quantitative definition of armchair and zigzag}
\begin{figure}[H]
    \centering
    \includegraphics[width=0.5\textwidth]{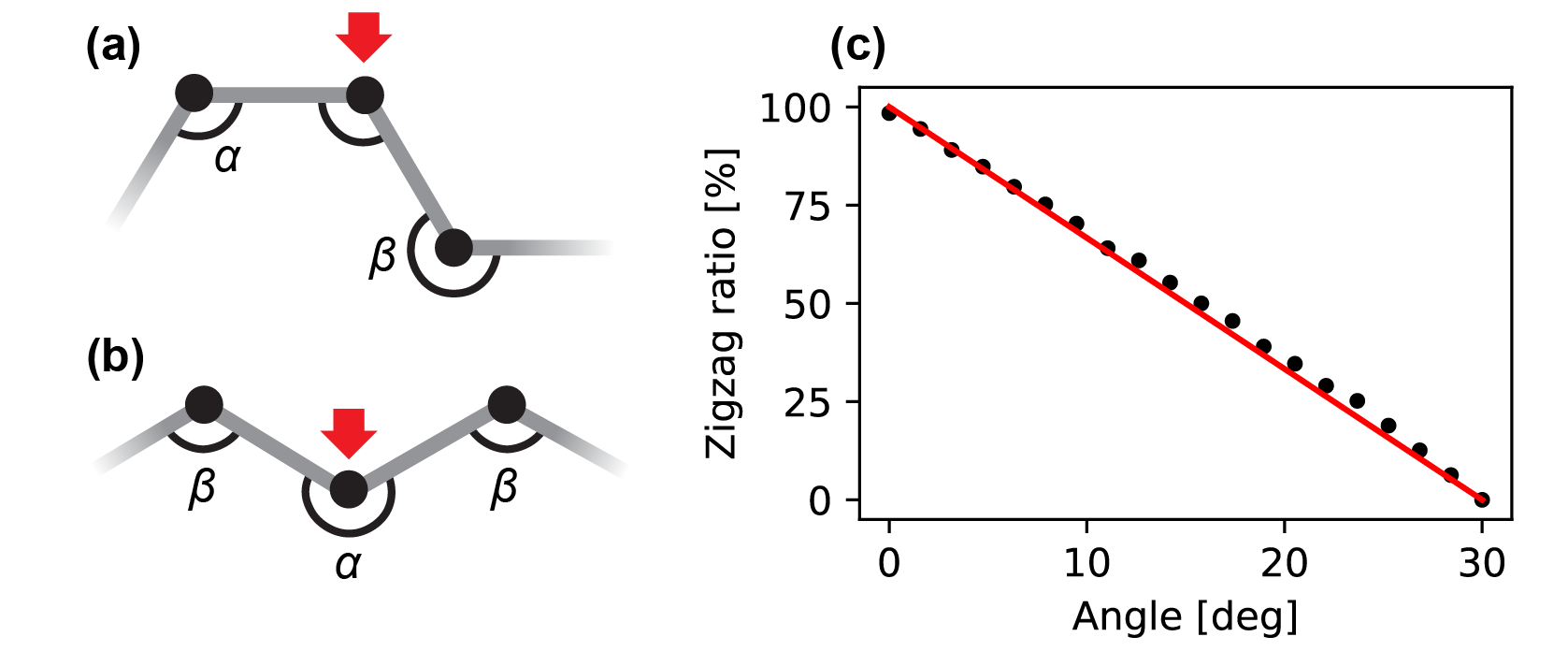}
    \caption{Definition of \emph{armchair vertex} and \emph{zigzag vertex} used in our paper. The vertex pointed by the red arrow is (a) an armchair vertex if its neighbours have signed angles different from each other, and (b) a zigzag vertex if its signed angle is different from its neighbours', and whose are equal to each other. (c) Ratio of zigzag vertices as a function of the angle of graphene interface.}
    \label{fig:vertex_types}
\end{figure}

As a measure of the amount of armchair and zigzag in the junctions, we introduced the concepts of armchair vertex and zigzag vertex, as shown in Fig. \ref{fig:vertex_types}(a) and (b). Note that these definitions are consistent when considering tilted graphene interfaces, where the relation between the angle $\alpha$ and the ratio of armchair and zigzags are proportional, as shown in Fig. \ref{fig:vertex_types}(c).
\section*{General considerations about the quality of the predictions}
\begin{figure}
    \centering
    \includegraphics[width=0.5\textwidth]{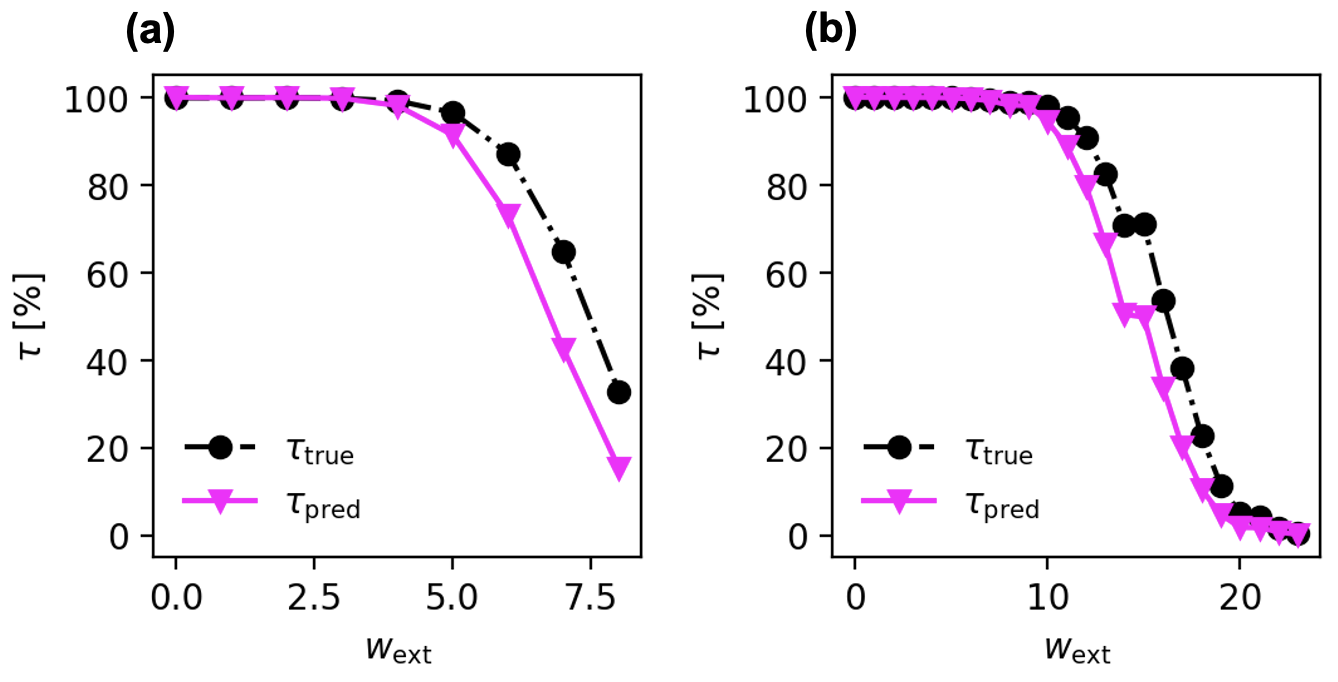}
    \caption{Preserved conductance $\tau$ (black dots) and predictions (magenta triangles) as function of the cutoff parameter $\wext$ (a) for 20-AGNR and (b) for 50-AGNR. Lines are only for readiability.}
    \label{fig:predictions}
\end{figure}

\begin{figure}
    \centering
    \includegraphics[width=0.5\textwidth]{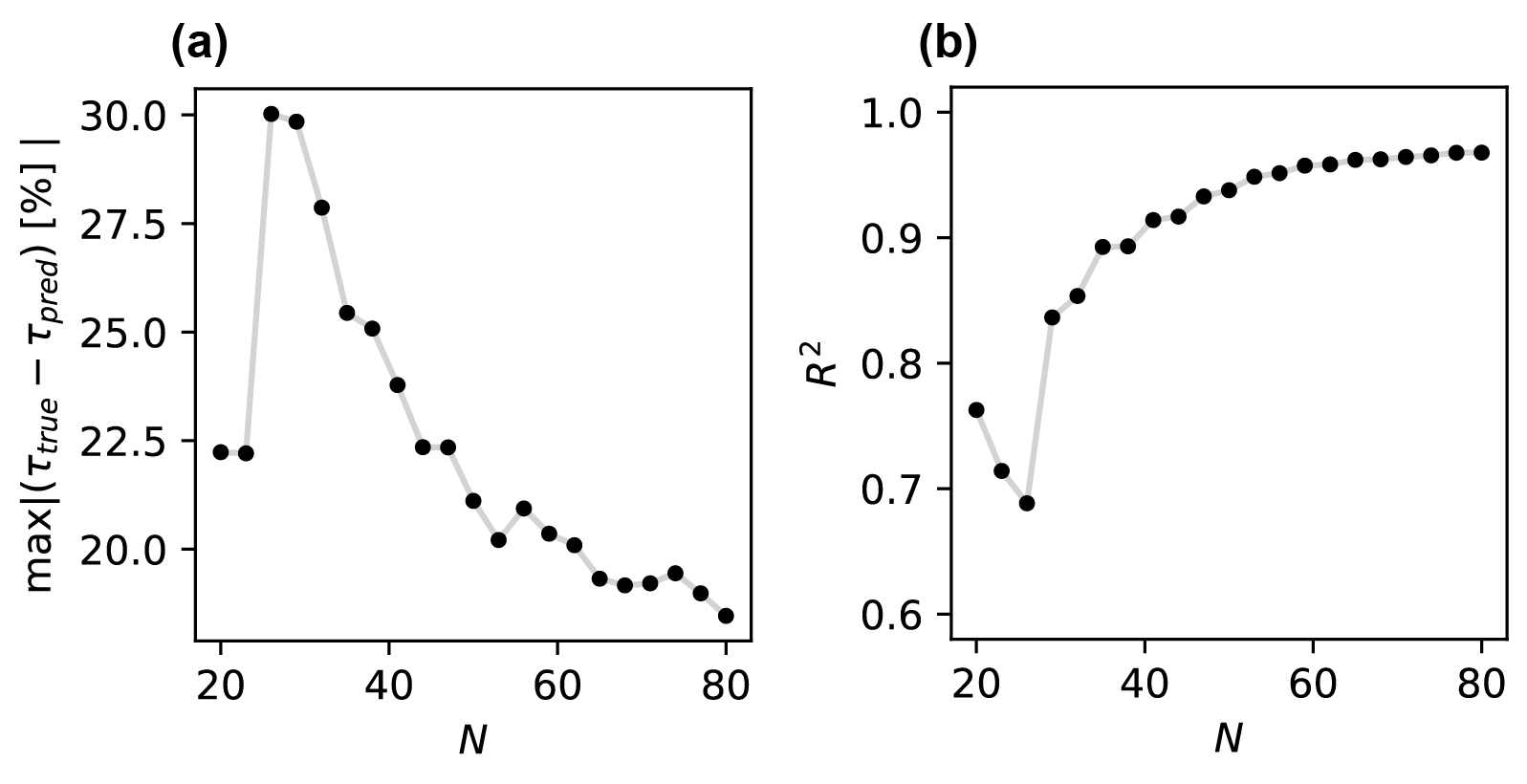}
    \caption{Quality of the preserved conductance predictions made using the geometric mean Ansatz at  $\delta E=0.00$ eV for different $N$. (a) Maximum error on the preserved conductance. (b) Coefficient of correlation $R^2$.}
    \label{fig:prediction_metrics}
\end{figure}

Fig. \ref{fig:predictions} shows more examples of predictions made using the geometric mean Ansatz.
Although the predictions both look of a similar quality, we can investigate the quality of the them using different metrics. One choice would be to consider the maximum error between the predictions $\tau_\text{pred}$ and the values obtained via simulations $\tau_\text{true}$. Looking at Fig.~\ref{fig:prediction_metrics}(a), it appears clearly that the error tends to diminish as we consider large AGNRs. Looking more globally at the quality of the fits, we can consider the coefficient of correlation $R^2$, as shown in Fig. \ref{fig:prediction_metrics}(b). Again, the larger the ribbons, the better the predictions.
\section*{More examples of rounded junctions}
\begin{figure}[h]
    \centering
    \includegraphics[width=.90\textwidth]{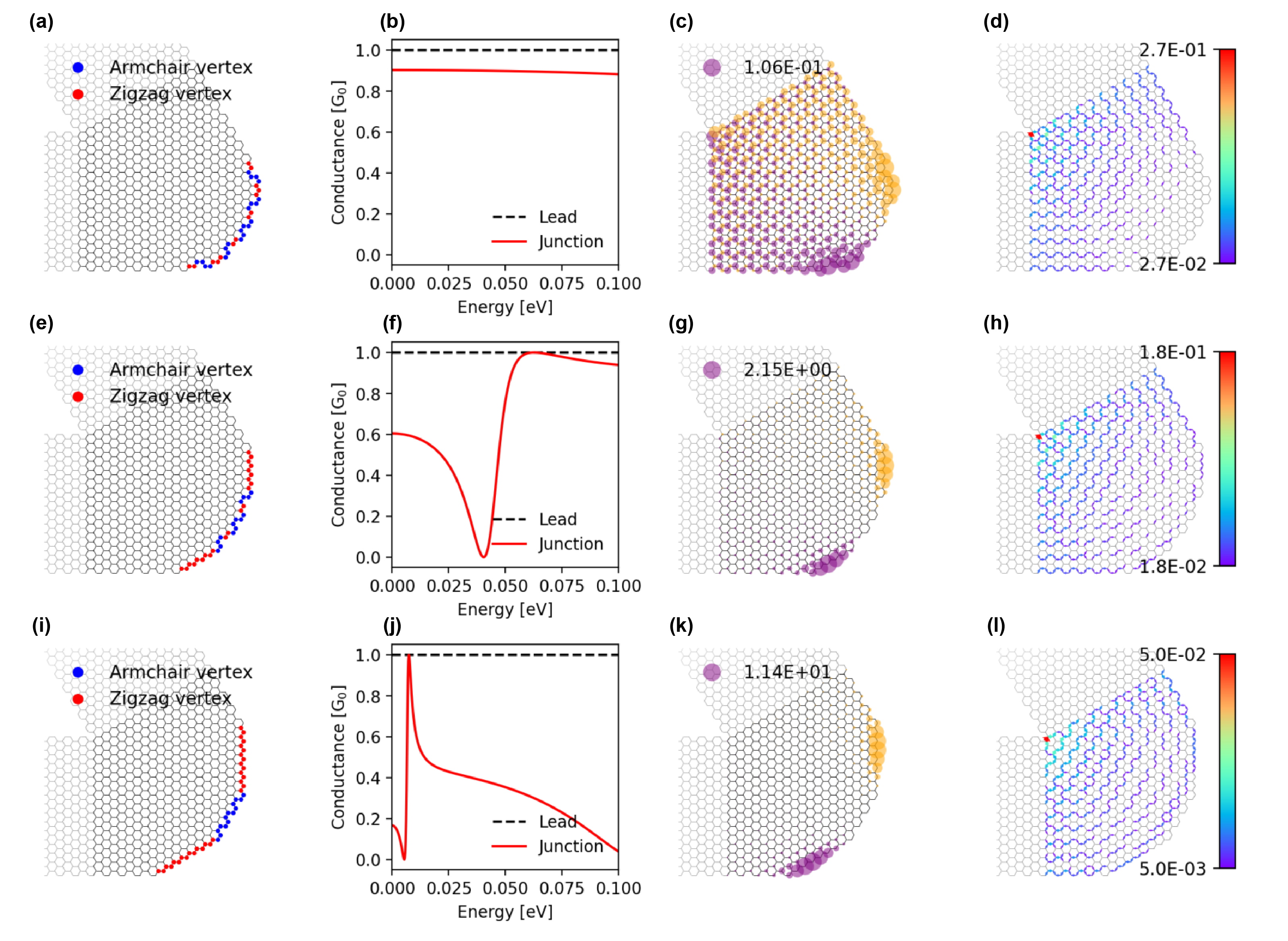}
    \caption{Edge structure, conductance, local density of states and probability currents of $32$-AGNR junctions  with a trimming (a)--(d) of $\wext=11$, (e)--(h) of $\wext=13$, and (i)--(l) of $\wext=15$.}
    \label{fig:varying_wext}
\end{figure}

\begin{figure}
    \centering
    \includegraphics[width=.90\textwidth]{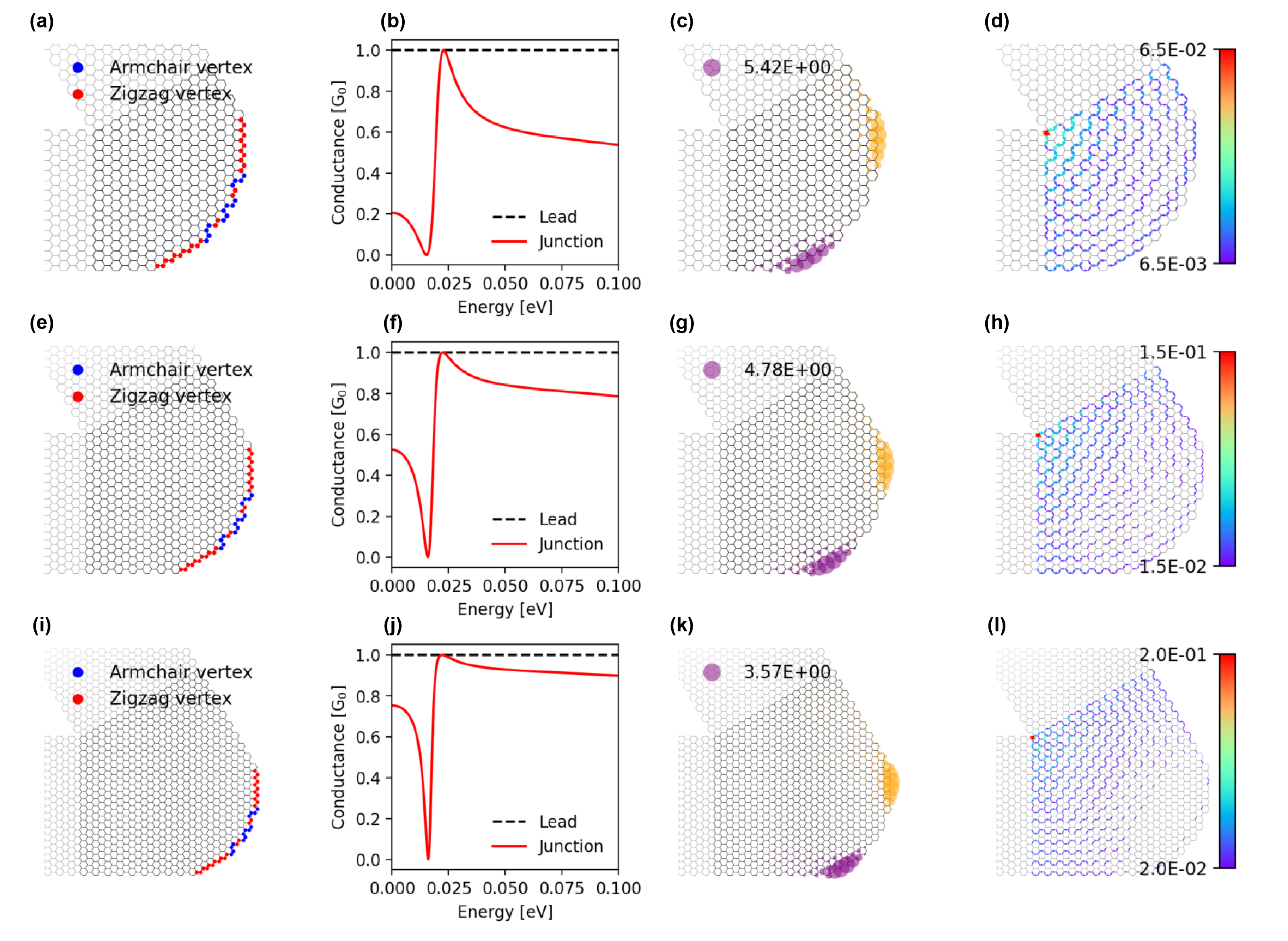}
    \caption{Edge structure, conductance, local density of states and probability currents of $N$-AGNR junctions  with a trimming of $\wext=12$ (a)--(d) with $N=29$, (e)--(h) with $N=35$, and (i)--(l) with $N=41$.}
    \label{fig:varying_N}
\end{figure}

Figs. \ref{fig:varying_wext} and \ref{fig:varying_N}  give more examples of junction's properties for various values of $\wext$ and $N$ respectively. Note that the complete database as well as all the numerical values can also be freely accessed online \cite{Leu23}.
\bibliography{references}